\documentclass[aps,pre,showpacs,preprint]{revtex4}
\usepackage{amscd}
\usepackage{graphicx}
\usepackage{bm}
\usepackage{amsmath,amssymb}

\begin{document}
\title{A new Monte Carlo method to study the fluid-solid phase transition of polydisperse hard spheres}

\author{ Mingcheng Yang}
\author{Hongru Ma}
\email{hrma@sjtu.edu.cn}

\affiliation{Institute of Theoretical Physics, Shanghai Jiao Tong
University, Shanghai 200240, People's Republic of China}
\date{\today}

\begin{abstract}

A new Monte Carlo approach is proposed to investigate the
fluid-solid phase transition of the polydisperse   system.
By using the extended ensemble, a reversible path was constructed to link
 the monodisperse and corresponding polydisperse system. Once
the fluid-solid coexistence  point of the monodisperse system is
known, the fluid-solid coexistence point of the polydisperse system
can be obtained from the simulation. The validity
of the method is checked  by the simulation of the fluid-solid phase
transition of a size-polydisperse hard sphere colloid. The results are
in agreement with the previous studies.

\end{abstract}

\pacs {05.10.Ln, 62.20.Dc, 82.70.Dd}

\maketitle


Melting and freezing transitions are among the most important
phenomena in condensed matter physics. The determination of the
equilibrium phase diagram of a specific material is the key point in
the understanding of the equilibrium properties of the material. The
hard sphere system(HSS), which is a representative model of a large
class of colloidal suspensions, has been extensively investigated.
It is well known that HSS undergoes an entropy driven first-order
phase transition from a disordered fluid to an ordered solid as the
packing fraction increases \cite{Hoover}. However, in a true
colloidal system the particles are inevitably  polydisperse in
particle sizes, which can be measured by the ratio of the standard
deviation to the mean of the diameter,
$\delta=\frac{\sqrt{\overline{(\sigma-\overline{\sigma})^{2}}}}
{\overline{\sigma}}$. It has a remarkable effect on the
thermodynamic and dynamic behaviors of the HSS
\cite{Pusey,Russel,Bolhuis,Fasolo,Martin, Schope}. The influence of
the polydispersity on the fluid-solid phase transition of the HSS
has been studied experimentally \cite{Pusey}, and theoretically by
computer simulation \cite{Bolhuis}, density functional theory
\cite{McRae,Chaudhuri}, moment free energy method \cite{Fasolo} and
other simplified theories \cite{Sear,Bartlett,Bartlett1}.
Experimentally, crystallization process depends on the
nonequilibrium effects \cite{Evans,Auer}, as a result the observed
phase behaviors deviate more or less from the equilibrium case.
However, the theoretically obtained phase diagram is usually the
equilibrium one. Therefore, the simulation study of freezing
transition of the polydisperse HSS is essential to check the
validity of the theory and to deeply understand the equilibrium
phase behaviors of the system.

When studying the fluid-solid coexistence of the polydisperse HSS by
computer simulations, there  exist two basic difficulties which also
exist in the monodisperse case: one is that the coexisting solid
forbids the particle insertion; and the other is that the back and
forth tunneling event between the liquid and solid phases never
happens within the simulation time scale. Therefore, some usual
methods, by which the phase transition of the polydisperse soft
spheres \cite{Wilding0,Fernandez} and the nematic-isotropic
transition of polydisperse liquid crystal \cite{Bates} can be
investigated effectively, are not suitable for the problem. To the
best of our knowledge, Gibbs-Duhem integration
algrithm\cite{Bolhuis} is the only available assumption-free method
to determine the phase behavior of polydisperse HSS, where the
coexistence curve is obtained by integrating the Clausius-Clapeyron
equation from the monodisperse hard sphere coexistence states.
However, the method is not very robust and lacks built-in
diagnostics, and no other simulation data are available for
comparison. Thus it is desirable to develop such an approach by which
the phase transition can be determined from the rigorous free energy
calculation. One possible scheme is to directly estimate the free
energy difference between the stable polydisperse hard sphere fluid
and solid phases. In principle, the semigrand ensemble lattice
switch method \cite{Yang,Wilding} is a feasible one. However, the
implementation of a simulation along this line is nontrivial. First,
the free energy difference between the fluid and the solid phases is
huge, second,  the system need to traverse a enormous entropy
barrier(``gate'' state), and third, the symmetry of fluid phase is
not the same as solid phase. The other possible scheme is the so
called referenced-state method \cite{Frenkel,Smit}. For the
polydisperse HSS the best reference  states are just the coexisting
fluid and solid of monodisperse HSS. Unfortunately, in the
simulation it is often difficult to find a reversible path linking
the monodisperse reference state to the polydisperse system under
consideration. The reason is that the ensemble for simulating the
polydisperse system(semigrand ensemble) is different from that for
the monodisperse system(canonical ensemble).

In the present work, we use the concept of extended ensemble to
construct a reversible path connecting the polydisperse target state
to the monodisperse reference  state, along which the free energy
difference between polydisperse and monodisperse system can be
estimated easily. By the method we reinvestigate the fluid-solid
phase transition of polydisperse HSS.

%

The semigrand canonical ensemble (SCE) is the best frame to
investigate the phase behavior of a polydisperse HSS. In this
ensemble the independent variable is the chemical potential
difference function, the total number of particles is fixed and the
number of particles of each size is permitted to fluctuate. A
continuous composition distribution can be realized on average.
Generally the composition distribution can not be prescribed in
advance unless we know the conjugated chemical potential difference
function. Fortunately, the chemical potential difference function
can easily be obtained by the recently developed SNEPR method, which
was described in detail in reference [\onlinecite{Yang, Wilding1}].
In all simulation studies of polydisperse systems the polydisperse
parameter(diameter of particles here) is always discretized. For the
purpose of describing our method more easily, we present the
semigrand canonical ensemble in the case of the discrete diameters.
The diameter of particles is assumed to take $M+1$ discrete values $
\sigma_{i}(\Delta)=\sigma_{m}+(i-\frac{M}{2})
\frac{\Delta}{M},
$ here $i=0$, $1$, $\cdots$, $M$,
$\Delta=\sigma_{max}-\sigma_{min}$, $\sigma_m
=(\sigma_{max}+\sigma_{min})/2$, $\sigma_{max}$ and $\sigma_{min}$
are the maximum and minimum of the allowed diameters of the
particles, respectively. In the limit of monodisperse case,
$\sigma_{max}=\sigma_{min}=\sigma_m$, $\Delta=0$, and all
$\sigma_i$'s are the same. However, we can still regard the
particles with different $i$'s as belonging to different species. We
will see that this concept is useful for later construction of
reversible paths between a monodisperse and a polydisperse system.

By introducing the excess chemical potential relative to ideal gas
$\mu_{ex}(\sigma_i)=\mu(\sigma_i)
-kT\ln(\frac{N\Lambda(\sigma_i)^{3}}{V})$, the partition function
$\Upsilon$ for the semigrand canonical ensemble is written as
\begin{eqnarray}
\Upsilon
=\frac{1}{N!\Lambda^{3N}(\sigma_{r})}\sum_{d_{1}=\sigma_0}^{\sigma_M}\cdots\sum_{d_{N}=\sigma_0}^{\sigma_M}
Z_{N}\times
\exp\left\{\beta\sum_{\alpha=1}^{N}(\mu_{ex}(d_{\alpha})-\mu_{ex}(\sigma_{r}))\right\}.\label{eq1}
\end{eqnarray}
Here, $\beta = 1/kT$, $\sigma_{r}$ is the diameter of an arbitrary
chosen referenced component, $\Lambda(\sigma_{r})=h/(2\pi
m_{r}kT)^{1/2}$ is the thermal wavelength of the referenced
component, $d_{\alpha}$ is the diameter of the $\alpha$th particle,
and $Z_{N}$ is the canonical configuration integral
\begin{equation}
Z_{N}=\int_{r_{1}}\cdots\int_{r_{N}}e^{-\beta
U}\prod_{\alpha=1}^{N}d\textbf{r}_{\alpha}.
\end{equation}
The partition function $\Upsilon$ is related to the semigrand
canonical free energy $Y$ through
\begin{equation}
Y=-kT\ln\Upsilon(N,V,T,\{\Delta\mu_{ex}(\sigma_i)\}),  \label{eq4}
\end{equation}
here $\Delta\mu_{ex}(\sigma_i)=\mu_{ex}(\sigma_i)-\mu_{ex}(\sigma_{r})$.


In order to establish a reversible path that links the polydisperse
system to the monodisperse referenced system, we employ an extended
semigrand canonical ensemble in the simulation. In the extended
ensemble the range of particle sizes
$\Delta=\sigma_{max}-\sigma_{min}$ is regarded as a new ensemble
variable. The extended ensemble is composed of the semigrand
canonical ensembles with different range of particle sizes $\Delta$.
Each value of $\Delta$ specifies a macroscopic state of the extended
ensemble. In what follows, it is convenient to set the diameter of
the referenced species $\sigma_{r}=\sigma_{m}$. With the additional
ensemble variable $\Delta$, the partition function of the extended
semigrand canonical ensemble is defined as
\begin{equation}
\Gamma(N,V,\{\Delta\mu_{ex}(\sigma_i)\})=\sum_{\{\Delta\}}
\Upsilon(N,V,\{\Delta\mu_{ex}(\sigma_i)\},\Delta),   \label{eq5}
\end{equation}
here $\Delta$ takes a series of discrete values ranged from
$\Delta=0$ to $\Delta=\Delta_{max}$, corresponding to the
monodisperse reference system and the polydisperse target system,
respectively. The $\{\Delta\mu_{ex}(\sigma_i)\}$ are kept the same
for all values of $\Delta$. For $\Delta=0$, the monodisperse case,
these $\{\Delta\mu_{ex}(\sigma_i)\}$ give a distribution of
particles among different species with the same diameters, while for
$\Delta=\Delta_{max}$ the $\{\Delta\mu_{ex}(\sigma_i)\}$ give the
required distribution of particles of different sizes. Thus, the
monodisperse system can be transformed into the polydisperse system
by  the variable $\Delta$. For the sake of convenience, the
mid-point of the particle sizes, $\sigma_m
=(\sigma_{max}+\sigma_{min})/2$, is fixed when changing $\Delta$. In
the simulation the $\Delta$ moves are performed by resizing all
particles, 
however, the partition
of the particles among species remains unchanged even though the
particle sizes are changed. 
The semigrand canonical free energy difference between
the macroscopic states $\Delta=\Delta_{max}$ and $\Delta=0$ is
written as
\begin{equation}
\Delta Y 
=\ln\left[\frac{Pr(0)}{Pr(\Delta_{max})}\right],  \label{eq6}
\end{equation}
where $Pr(\Delta)$ is the probability that the system is in the
macroscopic state $\Delta$. In the extended semigrand canonical
ensemble the probability can be obtained from simulation by the flat
histogram methods \cite{Berg,F.Wang,S.Wang}. In the simulation the
trial moves include particle translations, resizing particle
operations(breathing move) and changing $\Delta$ operations.

The semigrand canonical partition function with $\Delta = 0$ can be
written as
\begin{equation}
\Upsilon(N,V,\{\Delta\mu_{ex}(\sigma_i)\},0)
=\frac{Z_{N}}{N!\Lambda_m^{3N}} C(N,T,\{\Delta\mu_{ex}(\sigma_i)\})
 \label{eq7}
\end{equation}

Where the configuration integral $Z_{N}$ is independent of the
composition distribution when $\Delta=0$. And
$C(N,T,\{\Delta\mu_{ex}(\sigma_i)\})$ is a summation in the
particle species, independent of the volume. Equations (\ref{eq6})
and (\ref{eq7}) bridge between the semigrand canonical partition
function of polydisperse system and the canonical partition function
of monodisperse system.


As was well known, the fluid-solid phase transition of monodisperse
HSS can be determined by looking for the equal-weight double-peak
structure in the volume histogram\cite{Borgs}. In actual computation
we use a more tractable criteria ``equal peak height'' to locate the
transition point\cite{Challa}, which differs from equal-weight in
the finite-size effect. They give the same result in the
thermodynamic limit. If the number of particles of the coexisting
fluid is equal to that of the coexisting solid, according to the
``equal peak height'' condition their canonical partition functions
will satisfy the following relation
\begin{equation}
Z^{c}_{f} \texttt{exp}(-\beta P^{c}V^{c}_{f}) = Z^{c}_{s}
\texttt{exp}(-\beta P^{c}V^{c}_{s}), \label{eq8}
\end{equation}
here, $P^{c}$ is the coexistence pressure, $V^{c}_{f}$ and
$V^{c}_{s}$ are, respectively, the volume of the monodisperse
coexisting fluid and solid, and $Z^{c}_{f}$ and $Z^{c}_{s}$ are the
canonical partition functions of monodisperse coexisting fluid and
solid, respectively. Substituting (\ref{eq7}) in (\ref{eq8}), we get
\begin{equation}
\Upsilon_{f}(V^{c}_{f},\{\Delta\mu_{ex}(\sigma_i)\},0)
\texttt{exp}(-\beta P^{c}V^{c}_{f}) =
\Upsilon_{s}(V^{c}_{s},\{\Delta\mu_{ex}(\sigma_i)\},0)
\texttt{exp}(-\beta P^{c}V^{c}_{s}), \label{eq9}
\end{equation}
where $C(N,T,\{\Delta\mu_{ex}(\sigma_i)\})$ is eliminated for it
independent of the state of system. Thus, in the polydisperse system
with $\Delta=0$ the fluid phase is connected to the solid phase. For
the polydisperse system with $\Delta=\Delta_{max}$ that we consider,
equations (\ref{eq6}) and (\ref{eq9}) establish a relation between
the fluid phase and the solid phase.

\begin{figure}
\centering
\includegraphics[angle=0,width=0.9\textwidth]{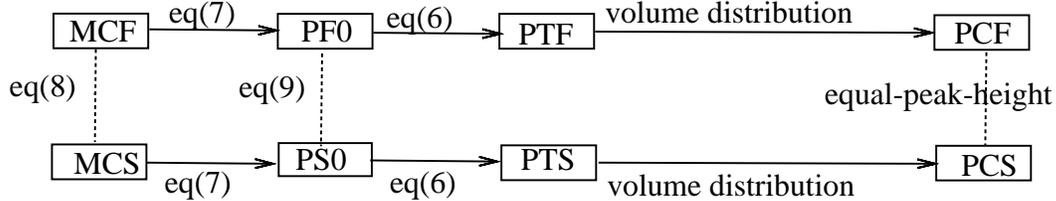}
\caption{The flow diagram of the simulation of polydisperse
fluid-solid coexistence. The acronym in the small boxes correspond
to: MCF and MCS: Monodisperse Coexisting Fluid and Solid; PF0 and
PS0: Polydisperse Fluid and Solid with $\Delta=0$; PTF and PTS:
Polydisperse Target Fluid and Solid with $\Delta=\Delta_{max}$; PCF
and PCS:  Polydisperse target Coexisting Fluid and Solid.  The
dotted lines describe the coexistence of  the fluid and solid
phases.}\label{fig1}
\end{figure}

Therefore, once the phase transition point of the monodisperse
system is located, combining equations (\ref{eq6}) and (\ref{eq9})
the semigrand canonical free energy difference between the solid and
fluid phases of the polydisperse target system with given
$\Delta_{max}$ and $\Delta\mu_{ex}(\sigma_i)$ can be calculated by
the extended semigrand canonical ensemble simulation where the
$\Delta$ is taken as the ensemble variable. In real calculations the
monodisperse coexistence point was taken from literature where the
fluid volume fraction is $0.492$ and the solid volume fraction is
$0.543$, respectively. The remaining task to locate the transition
point of the polydisperse target system is trivial. The polydisperse
target fluid and solid phases are treated as two new referenced
states, then the distribution of the volume of the polydisperse
target fluid and solid phases can be obtained separately by
performing another extended semigrand canonical ensemble simulation,
taking the volume $V$ as the ensemble variable. From such extended
ensemble simulations the $\Upsilon(V)$ can be calculated directly by
the flat histogram algorithm. When the volume distribution
$\Upsilon(V)\texttt{exp}(-\beta PV)$ shows two peaks of equal
height, the fluid-solid phases coexistence occurs. The flow diagram
of the calculation procedure is shown in figure \ref{fig1}.


\begin{figure}
\centering
\begin{minipage}[c]{0.4\textwidth}
\centering
\includegraphics[width=2.2in]{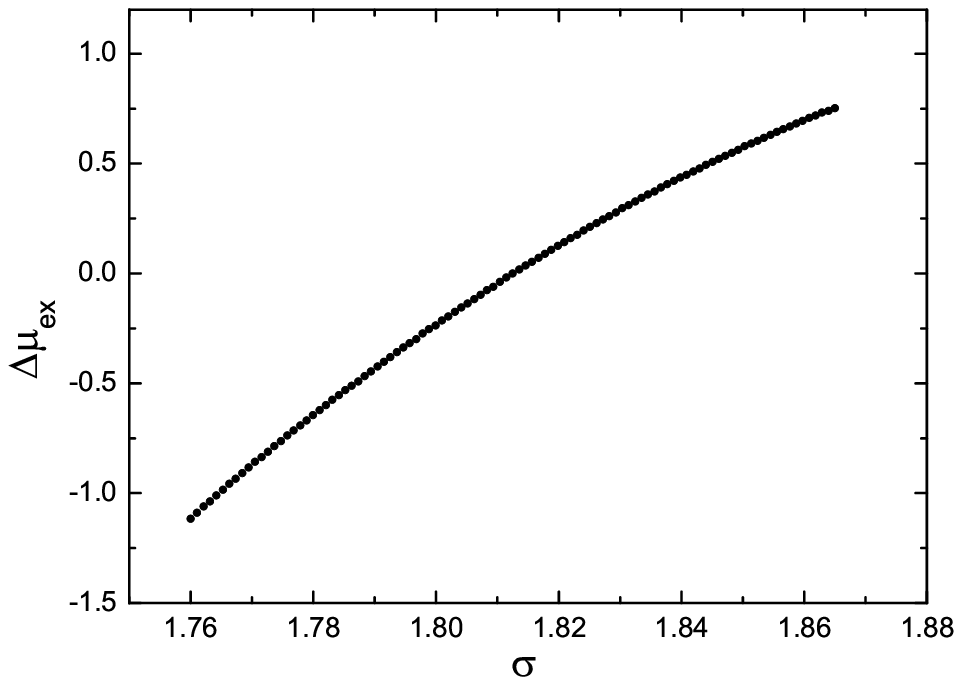}
\end{minipage}%
\begin{minipage}[c]{0.5\textwidth}
\centering
\includegraphics[width=2.8in]{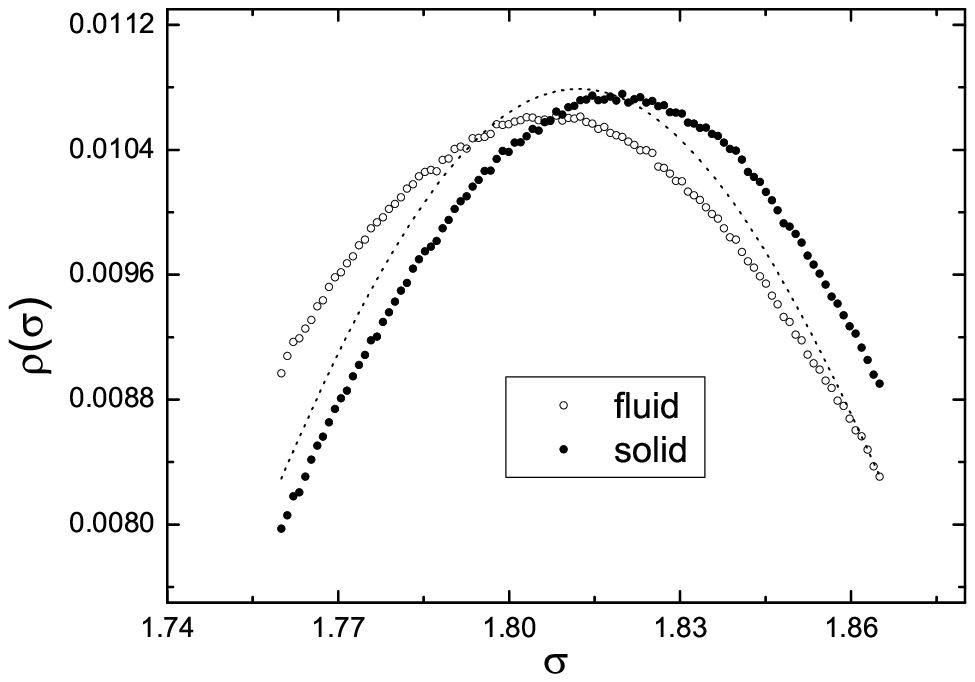}
\end{minipage}\\
\begin{minipage}[c]{0.4\textwidth}
\centering
\includegraphics[width=2.6in]{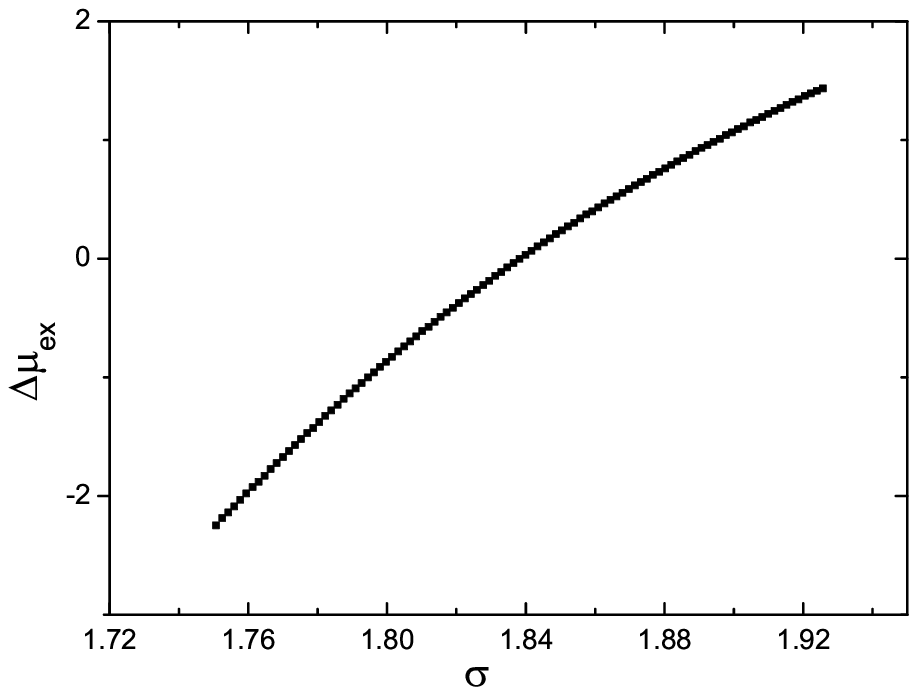}
\end{minipage}%
\begin{minipage}[c]{0.5\textwidth}
\centering
\includegraphics[width=2.8in]{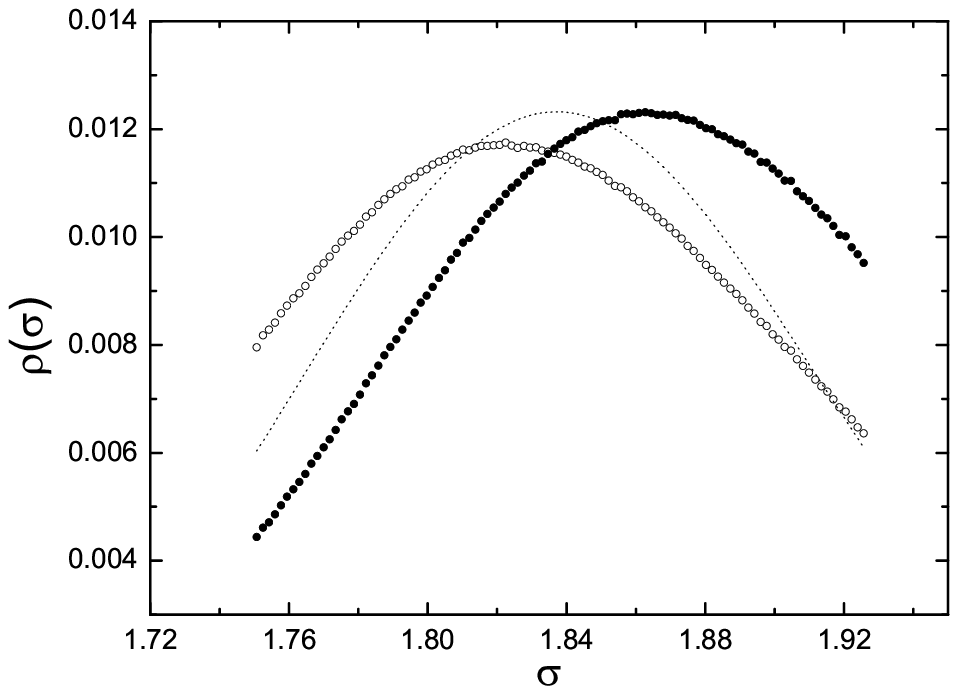}
\end{minipage}\\
\begin{minipage}[c]{0.4\textwidth}
\centering
\includegraphics[width=2.6in]{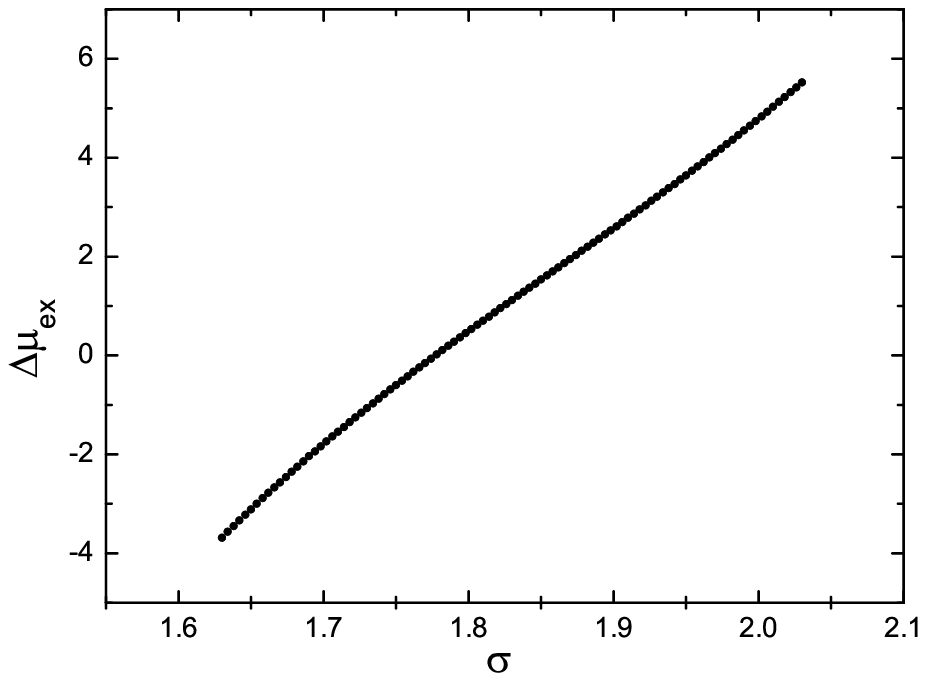}
\end{minipage}%
\begin{minipage}[c]{0.5\textwidth}
\centering
\includegraphics[width=2.8in]{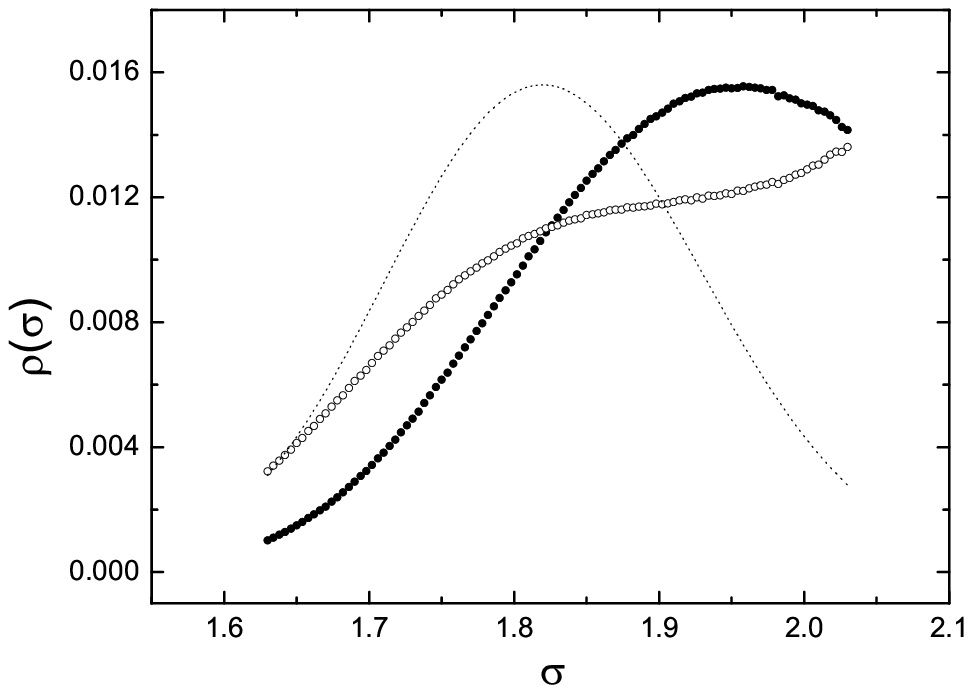}
\end{minipage}
\caption{Left: the chemical potential difference as a function of
the particle diameter. Right: the corresponding composition distributions
of the coexistence fluid phase(open circles) and solid phase(filled
circles). The dotted lines denote the prescribed composition
distribution of the initial target polydisperse crystal, here the
upper one with a size polydispersity of 0.0163, the middle one with
a size polydispersity of 0.0252, and the lower one with a size
polydispersity of 0.0507.}\label{fig2}
\end{figure}

To test the validity of our method, we use it to locate the
fluid-solid transition point of the polydisperse hard spheres. Our
system is composed of $256$ size polydisperse hard spheres contained
in a periodic box. The crystal structure under consideration is
face-center-cubic ($fcc$), since our previous calculation
\cite{Yang} showed that $fcc$ phase is still the most stable
structure of polydisperse hard sphere crystal. The composition
distribution of the initial target polydisperse solid linked to the
referenced monodisperse solid is the truncated Schultz function. The
conjugated chemical potential difference function is evaluated by
the SNEPR method. The solved chemical potential difference function
is saved and then used to perform the rest of the simulation. The
initial target fluid is produced by using the solved chemical
potential difference function. Of course, it is possible to
prescribe the composition distribution of the initial target fluid
rather than the initial target solid. Previous studies by Bolhuis et
al used the quadratic form for the chemical potential difference
function \cite{Bolhuis}, the corresponding composition distribution
is not the same as ours. However, for the same polydispersity
$\delta$ we expect that our simulation will give qualitatively
similar results.

\begin{figure}
\centering
\includegraphics[angle=0,width=0.45\textwidth]{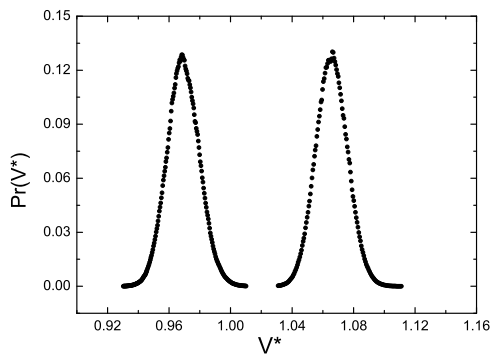}%
\caption{The probability distribution of the dimensionless volume of
the system at coexistence pressure, the imposed chemical potential
difference function is the upper one in Fig \ref{fig2}. The
dimensionless volume is defined as $V^{*}=V/N(\overline{\sigma})^3$,
here $\overline{\sigma}$ is the average   diameter of particles of
the initial polydisperse target crystal. }\label{fig3}
\end{figure}
\begin{figure}
\centering
\includegraphics[angle=0,width=0.45\textwidth]{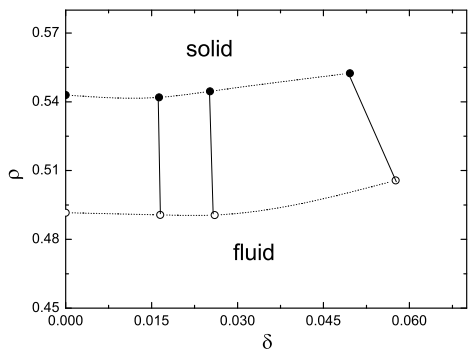}%
\caption{The phase diagram of the polydisperse hard sphere system in
the ($\delta$,$\rho$) plane. The dotted lines denote the boundaries of
the fluid and solid phases. The solid lines denote the tie lines
connecting the two coexisting phases.}\label{fig4}
\end{figure}
Using the SNEPR method we calculated the chemical potential
difference functions for three different truncated Schultz
distributions in the $fcc$ solid phases. Each distribution has
different polydispersity, and the maximum one is about
$\delta\simeq5\%$ because at higher $\delta$ the crystal may be
unstable \cite{Pusey,Bolhuis,Fasolo,Chaudhuri,Yang1}. Fig.\ref{fig2}
shows the chemical potential difference functions and the relevant
composition distributions of the coexisting phases. We see that the
particle size distribution of the coexisting fluid phase
significantly differs from that of the coexistence solid phase, this
is the phenomenon named as fractionation, the fractionation effect
increases with the polydispersity. Comparing with the coexisting
solid, the fluid has larger polydispersity and lower volume
fraction. These results are consistent with the previous results
\cite{Bolhuis,Fasolo1}. We also noted that the particle sizes
distribution of the coexisting solid differs from that of the
prescribed target solid, this is because the volume of initial
target solid is not equal to that of the coexisting solid, or we can
determine directly the cloud points(the coexisting phase with a
prescribed composition). Fig. \ref{fig3} is the probability
distribution of the volume at coexistence, the two equal height
peaks of the distribution correspond to the coexisting fluid and
solid phases, respectively. In Fig \ref{fig4}, we plot the phase
diagram of the polydisperse HSS in the ($\rho$,$\delta$) plane. For
the polydispersity used in the simulation, the phase diagram has the
same properties as the previous simulation \cite{Bolhuis}. The
transition points reported here are not accurately located because
extensive runs are needed to refine the locations, but it is suffice
to exhibit the validity of our method.


To conclude we provide a new Monte Carlo method to study the
fluid-solid phase transition of the polydisperse colloidal system.
The validity of the method is demonstrated by locating the
fluid-solid transition point of the polydisperse HHS. The obtained
results are consistent with the previous results. The method can be
directly extended to the polydisperse soft-interaction systems and
the hard-interaction system with other polydisperse attributes. It
presents a complement to the current Gibbs-Duhem integration method
\cite{Bolhuis}. The estimate of cloud points is important to fully
understand the phase behavior of polydisperse colloids
\cite{Buzzacchi}. It is possible to extend our current method to determine the cloud points
for the fluid-solid transition of a polydisperse system. We are working on this direction
and the results will be reported  in the near future.


The work is supported by the National Natural Science Foundation of
China under grant  No.10334020  and in part by the National Minister
of Education Program for Changjiang Scholars and Innovative Research
Team in University.


\end{document}